\documentclass[reprint,amsmath,amssymb,aps,prb]{revtex4-2}
\usepackage{graphicx}
\usepackage{color}
\usepackage{dcolumn}
\usepackage{latexsym}

\usepackage[normalem]{ulem}
\usepackage{hyperref,amssymb}
\usepackage{url}
\usepackage{color,soul}
\usepackage{graphicx}
\usepackage{verbatim}
\usepackage{multirow}
\usepackage{amsmath}
\newcommand{\beq}{\begin{eqnarray}}
\newcommand{\eeq}{\end{eqnarray}}
\usepackage{mathrsfs}
\usepackage{float,soul}
\usepackage[dvipsnames]{xcolor}
\usepackage{mathtools}
\usepackage{slashed}
\usepackage{physics}
\usepackage{graphicx}
\usepackage{epstopdf}
\usepackage{subfigure}
\usepackage[font=small]{caption}
\usepackage{hyperref}
\usepackage{bbold}
\usepackage{wasysym}
\usepackage{feynmp}
\usepackage{hyperref}
\hypersetup{colorlinks}
\usepackage{ragged2e}
\usepackage{siunitx}
\usepackage[utf8]{inputenc}

\begin{document}

\title{Many-body localization in a slowly varying potential}
\author{Zi-Jian Li}
\author{Yi-Ting Tu}
\author{Sankar Das Sarma}
\affiliation{Condensed Matter Theory Center and Joint Quantum Institute, Department of Physics, University of Maryland, College Park, Maryland 20742, USA}

\begin{abstract}
{We study many-body localization (MBL) in a nearest-neighbor hopping 1D lattice with a slowly varying (SV) on-site potential $U_j = \lambda\cos(\pi\alpha j^s)$ with $0<s<1$. The corresponding non-interacting 1D lattice model is known to have single-particle localization with mobility edges.
Using exact diagonalization, we find that the MBL of this model has similar features to the conventional MBL of extensively studied random or quasiperiodic (QP) models, including the transitions of eigenstate entanglement entropy (EE) and level statistics, and the logarithmic growth of EE. 
To further investigate the universal properties of this MBL transition in the asymptotic regime, we implement a real-space renormalization group (RG) method.
RG analysis shows a subvolume scaling $\sim L^{d_{\rm MBL}}$ with $d_{\rm MBL} \approx 1-s$ of the localization length (length of the largest thermal clusters) in this MBL phase. 
In addition, we explore the critical properties and find universal scalings of the EE and localization length. From these quantities, we compute the critical exponent $\nu$ using different parameters $s$ (characterizing different degrees of spatial variation of the imposed potential), finding the critical exponent staying around $\nu\approx2$. This exponent $\nu \approx 2$ is close to that of the QP model within the error bars but differs from the random model. This observation suggests that the SV and QP models may belong to the same universality class, which is, however, likely distinct from the random universality class. }
\end{abstract}
\maketitle
\section{Introduction}

Thermalization has destructive effects on preserving the information of quantum states, as most interacting systems reach thermal equilibrium, even when isolated from the environment. This interaction-induced self-thermalization or ergodicity is a fundamental pillar of statistical mechanics. 
Work over the last 20 years has, however, established that certain classes of 1D disordered interacting systems may escape thermalization even at infinite temperature by eigenstate localization, identified as the phenomenon of many-body localizations (MBL)~\cite{fleishman1980interactions,basko2006metal,oganesyan2007localization,nandkishore2015many,abanin2017recent,abanin2019manybody,alet2018many,sierant2025many,parameswaran2018many,imbrie2017local}. 
Theoretical investigations of MBL systems have demonstrated many unique features, including a logarithmic growth of the entanglement entropy (EE) after a quench~\cite{vznidarivc2008many,serbyn2013universal}, an area-law entanglement of all eigenstates~\cite{bauer2013area,serbyn2013local}, and an extensive set of local integral of motions (LIOMs) as an effective theoretical description~\cite{huse2014phenomenology}. 
Whether these MBL characteristics survive the infinite-time thermodynamic limit or not is still debated, but it is clear that for all practical purposes, the well-known eigenstate thermalization hypothesis may break down in real systems with large sizes for a long time~\cite{imbrie2016many}.

Understanding the critical behaviors of MBL transitions is limited by the finite-size effect in numerical simulations. To investigate the universal MBL properties, various real-space renormalization group (RG) schemes have been proposed~\cite{dumitrescu2017scaling, potter2015universal, vosk2015theory}. 
According to the RG, critical exponents of MBL transitions with quasi periodic (QP) disorders are different from those of random disorders~\cite{zhang2018universal}, suggesting the MBL transitions governed by randomness and quasi periodicity may belong to different universality classes. 
In general, random and QP MBL manifest rather similar qualitative phenomenologies except for the different numerical critical exponents.

Given the different critical exponents between random and QP models, an interesting question is whether there are still MBL universality classes with numerically different critical exponents (or perhaps with different qualitative MBL phenomenologies). 
At the single-particle level, there are other 1D models with localization transition properties (and the model characteristics) which are neither random nor QP. 
One such example is the deterministic model with a slowly varying (SV) potential~\cite{sarma1988mobility,sarma1990localization}, 
whose single-particle eigenstates can be exactly solved, 
manifesting a localization behavior distinct from both the corresponding non-interacting random and QP localization.
Although the single-particle SV model was extensively studied almost 40 years ago, the corresponding interacting model is rarely discussed in the MBL literature (except for some small-size numerical studies in Refs.~\cite{nag2017manybody,modak2018criterion}).

In the current work, we present an MBL study of the SV model in the presence of interaction, comparing its properties with MBL in random and QP models. In particular, we use the RG to evaluate the critical properties of SV MBL.
Specifically, we study an interacting SV model with the on-site potential $U_j = \lambda\cos(\pi\alpha j^s)$ (as in Refs.~\cite{sarma1988mobility,sarma1990localization}) with $0<s<1$, and establish the existence of an MBL transition and study its critical properties.
First, we provide the numerical evidence for the MBL through exact diagonalization (ED), by showing several features, 
including a transition in the EE, Wigner-Dyson to Possonian level statistics transition, and the logarithmic entanglement growth. 
Given the usual finite-size constraints ($L\sim 30$) associated with the exponential growth of the interacting Hilbert space in the ED methods, a real-space RG is implemented for systems up to $L=1000$ to study the general properties of the MBL in the SV model. 
Using the RG method in the MBL phase, we find the finite-size scaling of the localization length (length of the largest thermal clusters) satisfies a subvolume law $\xi_{\rm RG} \sim L^{d_{\rm MBL}}$ with $d_{\rm MBL}\approx 1-s$, which leads to approximately the same scaling behavior of the EE.
Due to the emergence of large clusters, we find that the MBL transition could be unstable when exceeding a particular system size, as reflected in the drift of the critical point to a larger potential strength. 
(This is most likely associated with the well-known, also in the random and QP models, feature of an upward drift in the critical disorder strength with increasing system size, implying an inherent fragility in MBL transitions.) 
In the regime where the MBL transition is stable, we find that the SV critical exponent is also robust with $\nu\approx2$ against the change of the parameter $s$. Then, by comparing this scaling exponent ($\approx 2$) to that of the QP model ($\nu=2.4\pm0.3$)~\cite{zhang2018universal}, we believe that the MBL transitions of the SV and QP model may belong to the same universality class, which is different from the randomness-induced MBL transitions, where $\nu \gtrsim 3$ is reported \cite{dumitrescu2017scaling, potter2015universal, vosk2015theory}.  
Given the numerical errors and the finite-size limitations of the calculations, we cannot rule out the possibility that the SV model (with $\nu \approx 2$) perhaps represents a still different universality MBL class from the QP model (with $\nu \approx 2$), but SV and QP do seem to manifest critical exponents well outside the error bars from the ones in the random MBL.

We mention that by ``MBL transitions'', we do not mean the rigorous MBL transitions in the conventional thermodynamic limit where the effect of avalanche instability is important~\cite{thiery2018many,morningstar2019renormalization,morningstar2022avalanches,sels2022bath,tu2023avalanche}, but rather an intermediate situation with an effective fixed point, as observed in previous studies of MBL transitions in random and QP models~\cite{vosk2015theory,potter2015universal,dumitrescu2017scaling,zhang2018universal}. 
Whether MBL exists as a thermodynamic transition remains an open question for the SV model as much as it does for random and QP models, but our work shows that MBL certainly exists as an effective transition in the SV model up to large length scales as much as it does in random and QP models. We comment that this effective MBL has been reported in many experimental systems \cite{schreiber2015observation,smith2016many,choi2016exploring,roushan2017spectroscopic,alex2019probing} even if the thermodynamic limit question remains unresolved.

This article is organized as follows. In Sec.~\ref{model}, we introduce the interacting SV model and review the several single-particle properties of the model. In Sec.~\ref{section_ED}, we study the properties of MBL transitions in the SV model using the ED method.
In Sec.~\ref{section_RGMBL}, we introduce the RG method and discuss the universal properties of the SV MBL transitions. 
In Sec.~\ref{section_Con}, we conclude with a summary and a discussion of possible future directions. In Appendix~\ref{apdxA}, we review the asymptotic solutions by providing some explicit results on the single-particle properties of the SV model. In Appendix~\ref{apdxB}, we discuss some properties of the interacting SV model, providing a comparison with random and QP MBL using ED.

\section{The model}
\label{model}
\begin{figure}
\centering
\includegraphics[width=0.9\linewidth]{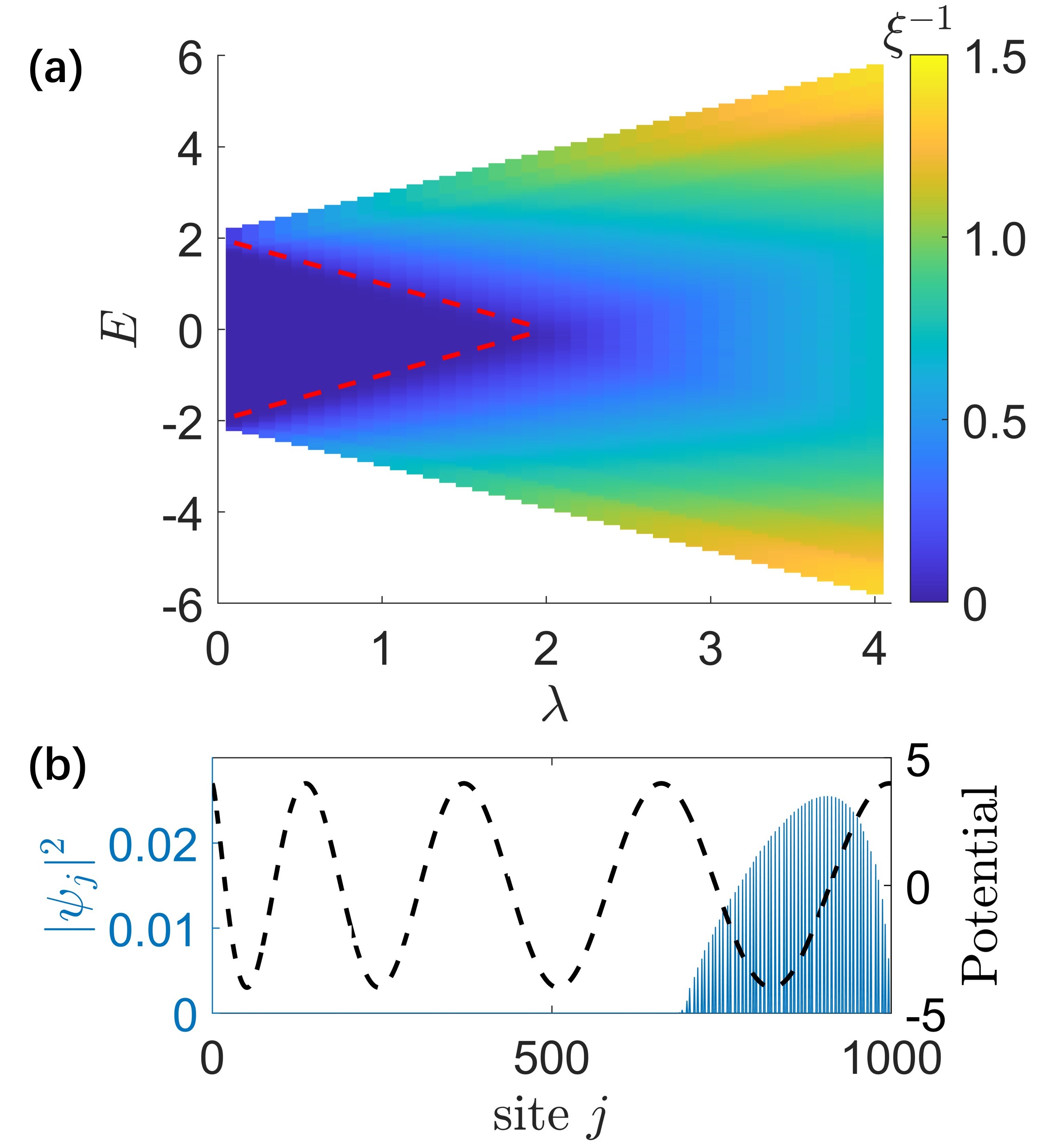}

\caption{\justifying {Single-particle properties ($V=0$) of the slowly varying model. (a) The inverse single-particle localization length of the slowly varying model with $\alpha=1,s=0.7,L=1000$. Red dashed lines are the analytical results for mobility edges $E_c = \pm (2-\lambda)$. (b) A typical localized eigenstate demonstrating the ``locally extended but globally localized'' property. The blue line shows the eigenstates and the black dashed line corresponds to the potential. Here, $\lambda=4, \pi \alpha=2,s=0.7,E=5.972$.}}
\label{SPL}
\end{figure}

We consider a 1D fermionic lattice with a SV deterministic potential~\cite{sarma1988mobility,sarma1990localization} with interactions
\begin{equation}
\begin{aligned}
    H_{\rm SV} = t \sum_{j=1}^{L-1} & (c_{j}^\dagger c_{j+1} + c_{j+1}^\dagger c_{j})\\ & + \lambda \sum_{j=1}^L \cos(\pi \alpha j^s + \phi) n_j 
    + V \sum_{j=1}^{L-1} n_j n_{j+1},
\end{aligned}
\label{H_SV}
\end{equation}
where $L$ is the length of the chain, $t$ is the hopping strength, $\lambda$ is the on-site potential strength, $V$ is the nearest-neighbor interaction strength. For the SV potential, the exponent is within $0<s<1$, the initial wavevector $\alpha$ has the order of 1, and the initial phase $\phi$ is averaged over. 
The corresponding single-particle model  (with $V=0$) was introduced and extensively studied analytically and numerically in Refs.~\cite{sarma1988mobility,sarma1990localization}, and we provide some details of the single-particle results in Appendix~\ref{apdxA}.
In the following discussions, the hopping strength is normalized to $t=1$ following the usual convention of the hopping energy being the unit of energy in the problem. 
The filling fraction is chosen to be $f=1/6$ in Sec.~\ref{section_ED}, but the results and conclusions should be independent of any specific choice of the filling factor. 
For $s=1$ and $\alpha$ irrational, the SV model becomes the well-known and extensively studied Aubry-Andr\'e (AA) QP model~\cite{aubry1980analyticity}. There are other existing QP models in the literature~\cite{biddle2009localization,biddle2010predicted,biddle2011localization,ganeshan2015nearest,li2017mobility,li2020mobility,vu2023generic} and comparisons have been made between our model and some of interacting QP models (i.e., Aubre-Andr\'e QP potential with next-neartest-neighbor hoppings studied in Ref.~\cite{zhang2018universal} in Sec.~\ref{critical} and Aubre-Andr\'e QP model in Appendix~\ref{apdxB}).

The single-particle properties of the SV model can be captured by the WKB approximation in the asymptotic limit~\cite{sarma1988mobility,sarma1990localization}. We briefly summarize the results of the single-particle SV model as follows. 
A single-particle mobility edge (SPME) exist at $E_c = \pm (2-\lambda)$ when $\lambda<2$, and the eigenstates within $\pm (2-\lambda)$ are extended. The eigenstates are fully localized when $\lambda>2$. Note that the exact values of $\alpha$, $s$, and $\phi$ are irrelevant to the SPME. The localized eigenstates in the SV model have a unique feature called ``locally extended but globally localized,'' where the localization length of the wave function averaged over particular regions around site $j$ diverges with $j$ as $\sim j^{1-s}/(s\alpha)$, but remains localized (i.e.\ much smaller than) compared with the size of the whole chain. A typical eigenstate and its background potential are shown in Fig.~\ref{SPL}(b). 
This characteristic ``locally extended but globally localized" feature indicates an unusual spatial inhomogeneity in the SV model arising from the slowly varying nature of the potential term, and makes the SV model different from random and QP models, which on average are spatially homogeneous on large length scales.
For detailed derivations of this property, see Appendix~\ref{apdxA}.

To demonstrate the localization properties, we compute the single-particle localization length of the $j$-th eigenstate using $\xi_{j}^{-1}  =\sum_{k \neq j} \ln |E_j-E_{k}| /(L-1) $~\cite{thouless1972relation}, where $E_j$ is the eigenenergy. For a fully localized state, the localization length approaches zero, while for fully extended states, it approaches infinity. We plot the inverse localization length in Fig.~\ref{SPL}(a) and mark the analytical result for mobility edges~\cite{sarma1988mobility,sarma1990localization}.

\section{Many-body localization from Exact Diagonalization}
\label{section_ED}
\begin{figure}
\centering
\includegraphics[width=1\linewidth]{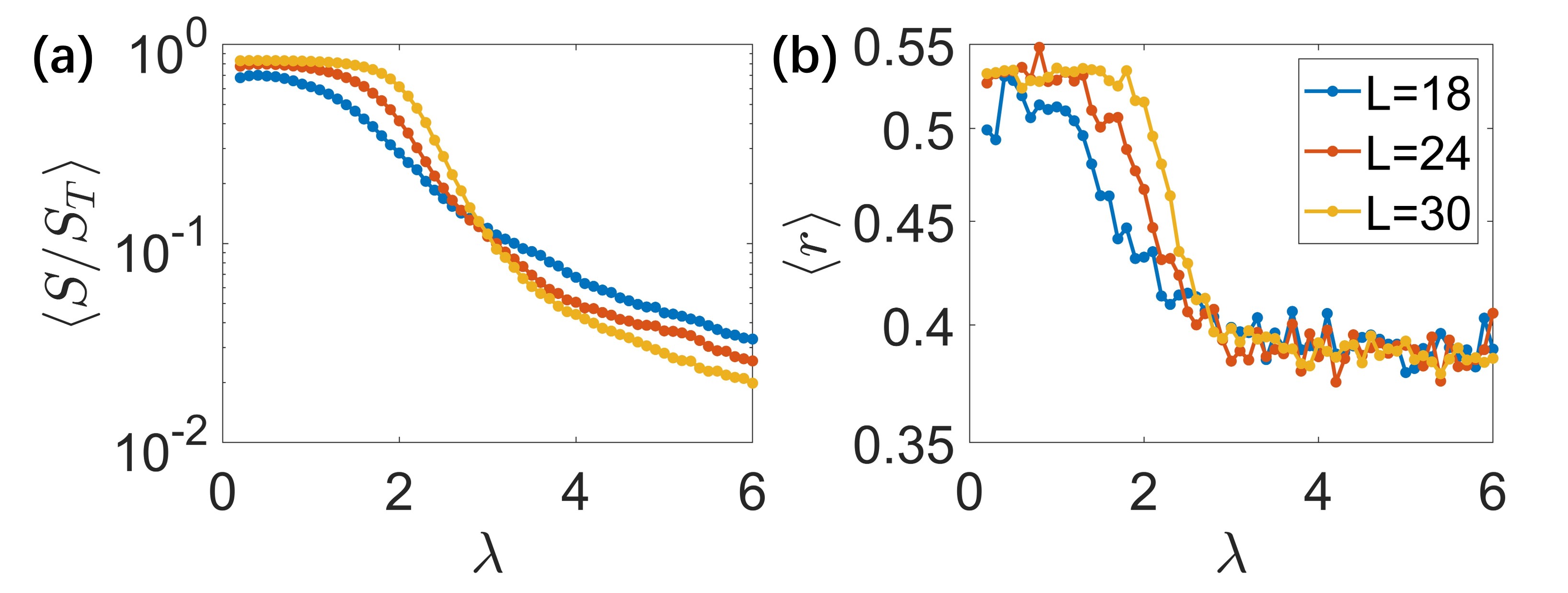}

\caption{\justifying {Entanglement entropy and the mean gap ratio of the slowly varying model. The parameters are $\alpha=1,s=0.7, V=1$ and filling fraction $f=1/6$. For $L=18,24$ and $30$, we average over the middle section of the spectrum with $100,10$ and $10$ initial phases, respectively. MBL transitions are apparent in both figures around a critical disorder strength $\lambda_c\sim 3$ where the individual curves of different system sizes cross through an approximately single disorder strength. }}
\label{ED}
\end{figure}

To identify the possible MBL transition in the model, we calculate the EE and the mean gap ratio at different $\lambda$. The EE of a subsystem $A$ is defined as $S(l) = - {\rm Tr} \rho_A \ln \rho_A$, where $\rho_A = \Tr_B \rho$ is the reduced density matrix of the subsystem $A$, ranging from site $1$ to site $l$. We choose $l=L/2$ throughout. The maximum (thermal) EE is given by the Page value $S_T = (H(f)L-1)/2$~\cite{page1993average}, where $H(f) = -f\ln(f)-(1-f)\ln(1-f)$ with the filling fraction $f$.
In the thermal phase, the EE satisfies the volume law ($S\sim L^d$ where $d=1$ is the dimension of the system), reaching the Page value, where $S/S_T\rightarrow 1$.
In the MBL phase, it usually follows the area law ($S\sim L^{d-1}$), so in the thermodynamic limit $S/S_T\rightarrow 0$. 

We use the level statistics analysis to identify the non-thermal behavior in the MBL phase. For the many-body spectra, the gap ratio is $r_n = \min(\delta E_n,\delta E_{n+1})/\max(\delta E_n,\delta E_{n+1})$,
where $\delta E_n = E_{n+1} - E_{n}$ is the difference between two adjacent eigenvalues. The gap ratio of the thermal phase follows the Wigner-Dyson distribution, whose mean value is $\langle r \rangle = 0.53$. In the non-thermal phase (i.e., the MBL phase), it obeys Poisson distribution, and the mean value is $\langle r \rangle = 0.38$.

In Figs.~\ref{ED}(a) and \ref{ED}(b), we provide the ED results for the EE and level statistics, respectively. Both figures indicate a thermal-to-MBL transition around a critical disorder strength $\lambda_c\approx 3$, which is larger than the non-interacting critical point at $\lambda=2$ as expected (and is comparable to the critical disorder strength for MBL transitions in both random and QP models, although the meaning of disorder in different models is quite different).
Note that although the EE in the localized phase appears to follow the area law, in the next section we will explore its asymptotic behavior, which should behave as a subvolume law $S\sim L^{d_{\rm MBL}}$ with $d_{\rm MBL}\approx 1-s$.
In our parameter regime where carrying out ED is computationally feasible, this behavior is almost indistinguishable from the area law (and we continue referring to this behavior as the area law in the current context).

\begin{figure}
\centering
\includegraphics[width=\linewidth]{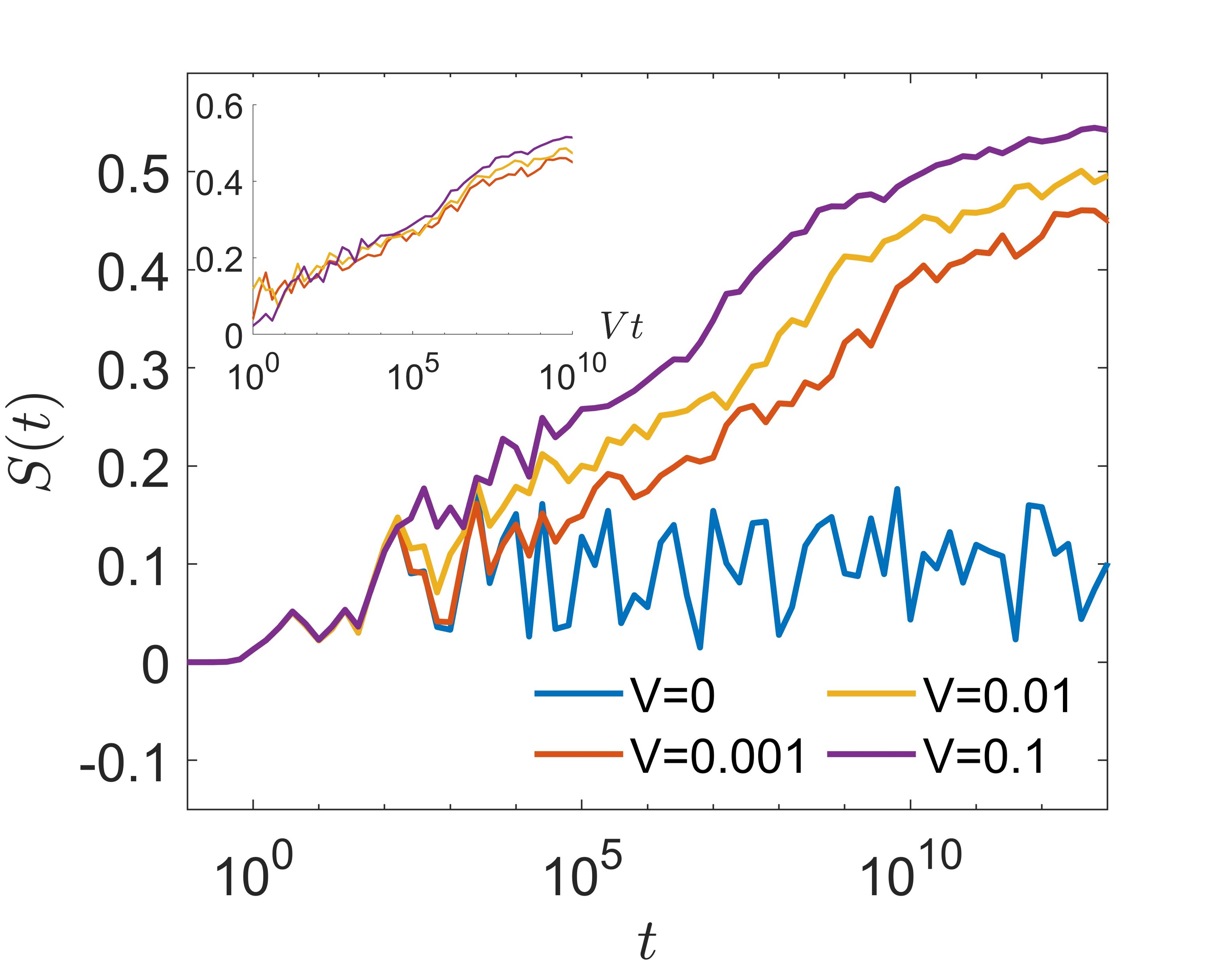}

\caption{\justifying {Averaged entanglement entropy growth of the slowly varying model. The potential strength is $\lambda=5$. We fix $s=0.7$, $\alpha=1$, and $\phi=0$. Saturation of the $V=0$ noninteracting results at short times indicates diffusive transport. The inset shows the logarithmic growth of interacting results for $Vt \gtrsim 1$. }}
\label{S_t}
\end{figure}

To provide more evidence for the SV MBL phase, we compute the EE growth from initial weakly entangled states. 
In Fig.~\ref{S_t}, we randomly choose $1000$ product states where the particles are not filled into the nearest and next-nearest-neighbor sites of the cut as initial states and compute the time evolution of the bipartite EE (averaged over these product states). 
For the non-interacting model, the EE saturates to a value at a relatively short timescale, due to the fast diffusive transport. 
In the presence of interactions, the EE continues to grow logarithmically for the timescale $Vt\gtrsim 1$ (inset of Fig.~\ref{S_t}), 
which is additional evidence for the MBL phase. Other properties of the interacting SV model and its comparison with other models using ED are provided in Appendix~\ref{apdxB}.

The properties of the SV MBL phase may be significantly different when $L$ is large. 
To gain some intuition, we consider the non-interacting model. As mentioned in Sec.~\ref{model}, the single-particle eigenstates are locally extended in the asymptotic regime. 
By adding interactions, these locally extended regions may effectively thermalize to large clusters. 
Due to the size limitations in the ED, these clusters cannot be probed directly numerically. Therefore, to study the general properties of the SV MBL transitions, a real-space RG method is used in the next section. 
We do, however, emphasize that we have established in the current section the existence of the MBL transition in the interacting SV model, as well as it is established in the literature within the ED technique in random and QP models.

\section{Renormalization group and MBL transitions}
\label{section_RGMBL}
To study the general properties of the SV MBL transition, we first review the real-space RG used in Refs.~\cite {dumitrescu2017scaling,zhang2018universal} for studying the MBL in random and QP models. 
The intuition of the RG is explained as follows using an effective ``spin" system as an example. The nature of the MBL transition can be understood as the decreasing resonances of distant spins. 
Such a process is governed by two energy scales---the tunneling amplitude $\Gamma_{ij}$ and the energy mismatch $\Delta E_{ij}$, where $i(j)$ denotes the $i(j)$-th spin. When $\Gamma_{ij}\gg\Delta E_{ij}$, the local spins $i,j$ become resonant, and the information spreading between different states is fast (i.e., diffusive). On the other hand, if $\Gamma_{ij}\ll\Delta E_{ij}$, the direct spreading of information from spin $i$ to spin $j$ is nearly impossible. In the RG scheme, all the resonant spins are combined into several different clusters. In the vicinity of the MBL transition, the clusters should occur in a scale-invariant way, leading to an RG scenario in real space. In practice, we treat these ``spins" as the single-particle degrees of freedom (i.e., the single-particle eigenstates).
The real-space RG here is somewhat of a heuristic, but the technique allows numerics on large system sizes, thus circumventing the severe finite-size constraint of the ED technique. 
This technique is useful in determining the MBL critical properties in large systems. 

The RG here focuses on the structure of the cluster $i'$, which consists of a set of single-particle degrees of freedom that are resonant. The clusters are characterized by an additional scaling parameter $\Lambda_{i'}$ and the number of the single-particle degrees of freedom $n_{i'}$. 
The many-body level spacing of a cluster is defined as $\delta_{i'} = \Lambda_{i'}/(2^{n_{i'}}-1)$, which are used to obtain the energy mismatch $\Delta E_{i'j'}$. The tunneling amplitude between clusters $\Gamma_{i'j'}$ and the energy mismatch $\Delta E_{i'j'}$ are used to characterize the merging process of different clusters in the RG flow, determining criticality. 

When starting the RG flow, each single-particle eigenstate with eigenenergy $\varepsilon_i$ is considered as a single cluster ($n_i$=1) with $\Lambda_{i} = \varepsilon_i$. The parameters of the clusters are initialized according to the single-particle properties with $\Delta E_{ij} = |\varepsilon_i - \varepsilon_j|$ and $\Gamma_{ij} = V\exp[-|r(i)-r(j)|/\max(\xi_i,\xi_j)]$, where $r(i)$ is the charge center defined by the expectation value of the position of the eigenstate, and $\xi_i$ is the localization length of the single-particle eigenstate. We set $V=0.5$ throughout. 

The RG flow begins after the initialization. We introduce a dimensionless parameter $g_{ij} = \Gamma_{ij}/\Delta E_{ij}$ to characterize whether two clusters are thermal or localized. If $g_{ij}>1$, a resonant bond is assigned between the clusters $i$ and $j$. Clusters connected by a path of resonant bonds are merged into a new cluster $\{i\} \rightarrow I$ containing $n_I = \sum_{i\in I} n_i$ single-particle degrees of freedom. Then the parameters of the new cluster $I$ are renormalized as 
\begin{equation}
    \Lambda_{I} = \sqrt{\sum_{i\in I} \Lambda_i^2 + \sum_{i,j\in I} \Gamma_{ij}^2}
\end{equation}
with $\delta_I = \Lambda_{I}/(2^{n_I}-1)$ by definition. 

After obtaining all the new clusters, we update their tunneling amplitudes and energy mismatches. The tunneling amplitude between two clusters $I$ and $J$ is 
\begin{equation}
    \Gamma_{IJ} = \left(\max_{i \in I,j \in J} \Gamma_{ij}\right) e^{-(n_{I}+n_{J}-n_{i}-n_{j})s_{\rm th}/2},
\end{equation}
where $\Gamma_{ij}$ denotes the tunneling of the clusters in the last step and $s_{\rm th}=\ln 2$ is the thermal entropy density. If two clusters are both unchanged in the same step, the couplings are turned off $\Gamma_{IJ} = 0$, and these clusters then become localized.

The energy mismatch is updated as 
\begin{equation}
    \Delta E_{IJ} = \left\{
    \begin{array}{ll}
        \max(\delta_I - \Lambda_J,\delta_J), & \Lambda_I \geq \delta_I \geq \Lambda_J \geq \delta_J, \\
        \max(\delta_J - \Lambda_I,\delta_I), & \Lambda_J \geq \delta_J \geq \Lambda_I \geq \delta_I, \\
        \delta_{I}\delta_{J}/\min(\Lambda_{I},\Lambda_{J}), & {\rm otherwise.}
    \end{array}
    \right.
\end{equation}

RG steps are iterated until no additional resonant bonds are added or the whole system has thermalized into a single cluster, thus completing the RG flow. 

We use two quantities to characterize the general properties of the MBL transition after the RG flow converges and stops~\cite{potter2015universal,dumitrescu2017scaling,zhang2018universal}. 
In particular, we study the normalized EE $s(x) = \min (m,n)/x$, where the cut $x$ partitions the cluster into $m$ and $n$ sites. The cut is set to $x=L/2$ throughout. 
In the thermal phase, the clusters are as large as the system size with a probability close to $1$, therefore, the normalized EE always approaches unity. 
Obviously, the normalized EE vanishes in the MBL phase. 
In addition, we also use the many-body localization length $\xi_{\rm RG}$, defined as the $n_i$ of the largest cluster. In the thermal phase, the localization length equals the system size $\xi_{\rm RG}\approx L$. The quantity $\xi_{\rm RG}/L$ crosses over from $1$ to $0$ as the system transitions from the thermal phase to the localized phase by varying the potential strength $\lambda$, which is the standard disorder tuning parameter controlling MBL.

\subsection{Structure of clusters in the MBL phase}

\label{section_RG}
\begin{figure}
\centering
\includegraphics[width=\linewidth]{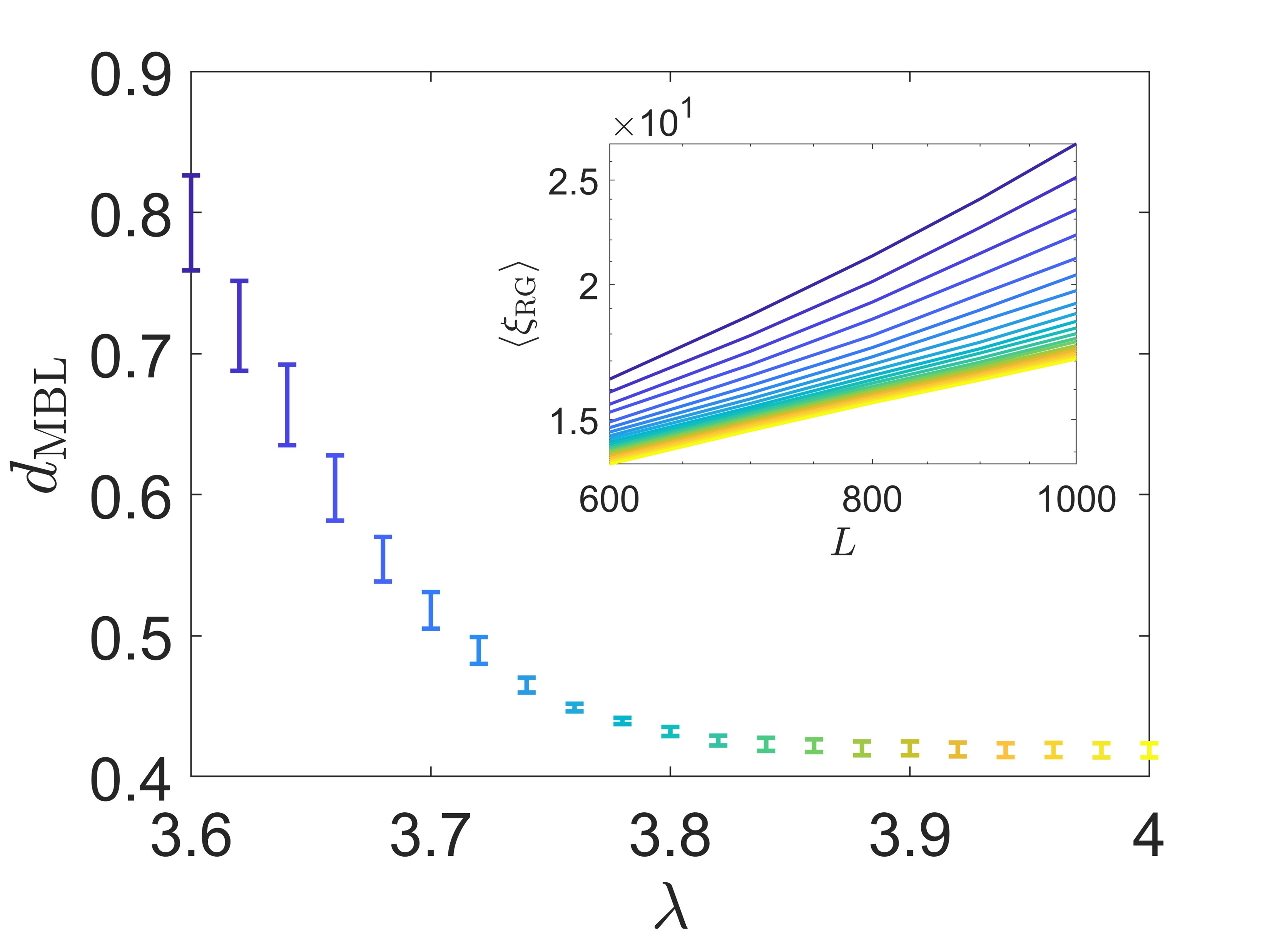}

\caption{\justifying {The saturation of the finite-size scaling exponent $d_{\rm MBL}$ of the largest cluster for large $\lambda$ (in the MBL phase). The parameters are $\alpha=1,s=0.6$. The inset shows the original data that are used to extract the exponent. The exponent approaches a value slightly larger than $1-s$ for large $\lambda$. }}
\label{nu_MBL}
\end{figure}

\label{section_RG}
\begin{figure}
\centering
\includegraphics[width=\linewidth]{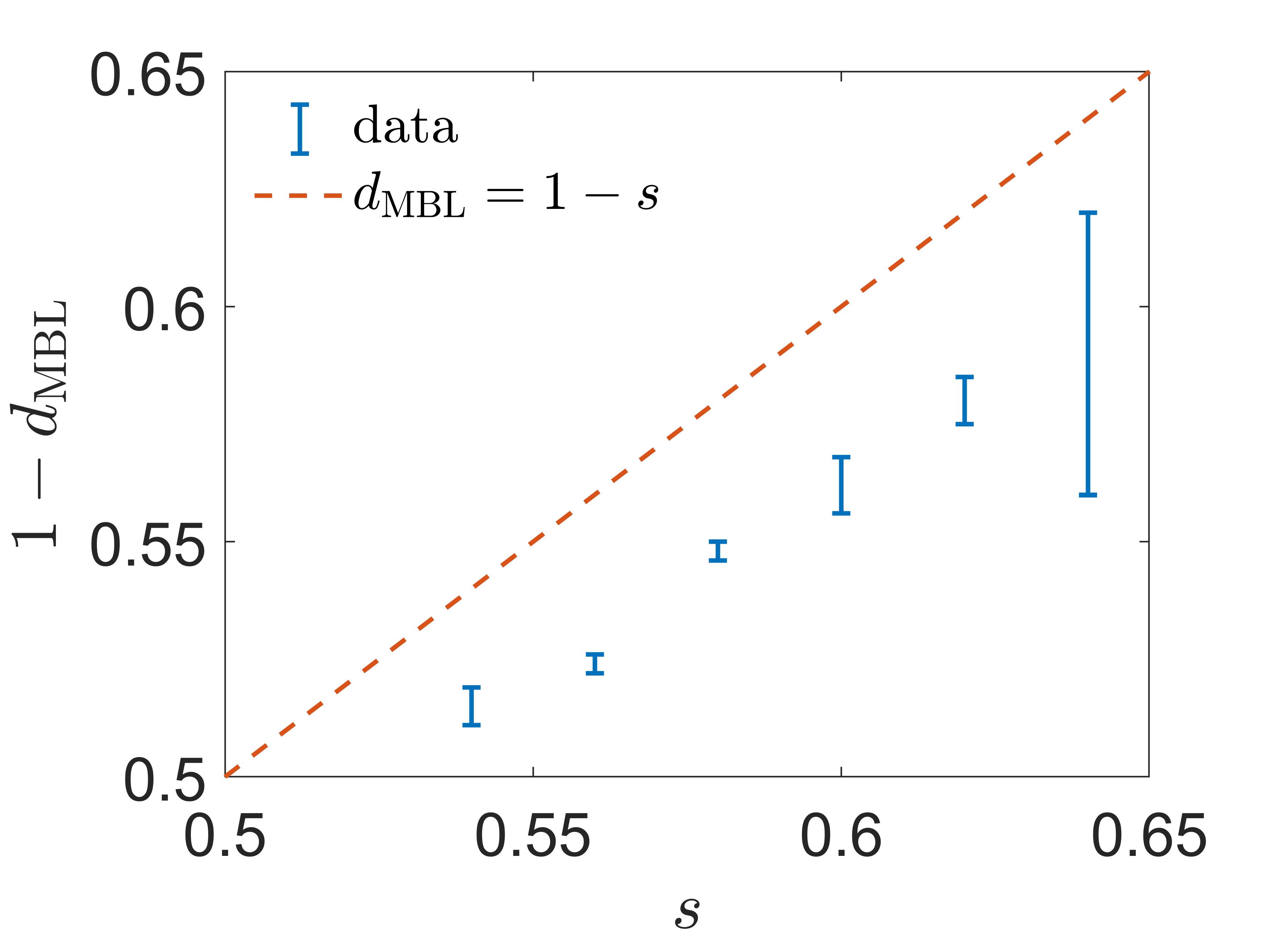}

\caption{\justifying {Saturated finite-size scaling exponent $d_{\rm MBL}$ as a function of the slowly varying potential exponent $s$ in the MBL phase.}}
\label{d_MBL_s}
\end{figure}

Before discussing the critical properties of the SV MBL transitions, we first study the structure of the RG clusters in the MBL phase. 
As previously mentioned, the locally extended regions of the non-interacting model may thermalize into local ergodic clusters in the presence of the interactions, the largest of which has a length diverging with the system size. To see this, we consider the localization length $\xi_{\rm RG}$ at sufficiently large $\lambda$, that is, in the MBL phase. In conventional MBL systems (e.g., the random Anderson disorder model), $\xi_{\rm RG}$ is usually a constant in the MBL phase, reflecting the LIOM structure. 
By contrast, in the SV MBL phase we are considering, $\xi_{\rm RG}$ scales with the system size as a subvolume law $\xi_{\rm RG}\sim L^{d_{\rm MBL}}$ with $0<d_{\rm MBL}<1$. 
Using the real-space RG, we find that the scaling exponent $d_{\rm MBL}$ saturates to a fixed value, related to the SV exponent $d_{\rm MBL} \approx 1-s$, for large potential strength (Fig.~\ref{nu_MBL}). 
Note that the largest locally extended region in the single-particle localized states has approximately the same scaling relation as in the MBL situation. This confirms the intuition that the locally extended regions in the absence of interactions thermalize to ergodic clusters in the presence of interactions. As a verification, we compute the scaling exponent $d_{\rm MBL}$ as a function of the SV exponents $s$ in Fig.~\ref{d_MBL_s}. 
The saturated value of $d_{\rm MBL}$ shows a linear dependence on $s$. Note that $1-d_{\rm MBL}$ is slightly smaller than $s$, possibly due to interaction (and/or finite-size) effects. 

\begin{figure*}
\centering
\includegraphics[width=1\linewidth]{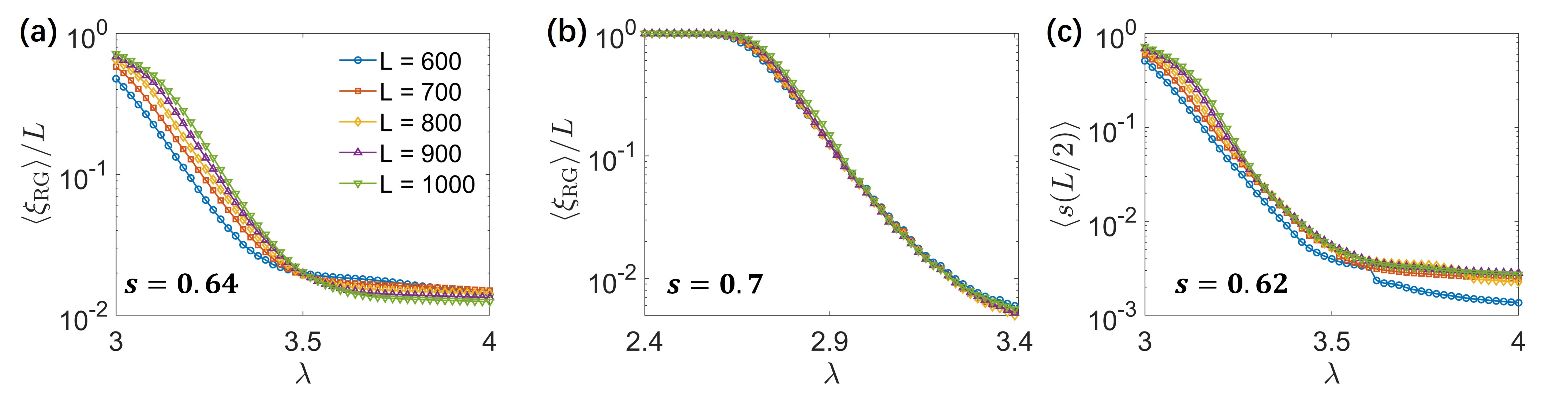}

\caption{\justifying {The failure of the RG around $s\gtrsim0.62$ below a particular system size. Normalized localization length from RG near criticality for (a) $s=0.64$ and (b) $0.7$. (c) Normalized EE from RG for $s=0.7$. For all figures, $\alpha=1$. The curves are averaged over $\sim 10^5$ initial phases.}}
\label{RG_fail}
\end{figure*}

\begin{figure}
\centering
\includegraphics[width=1\linewidth]{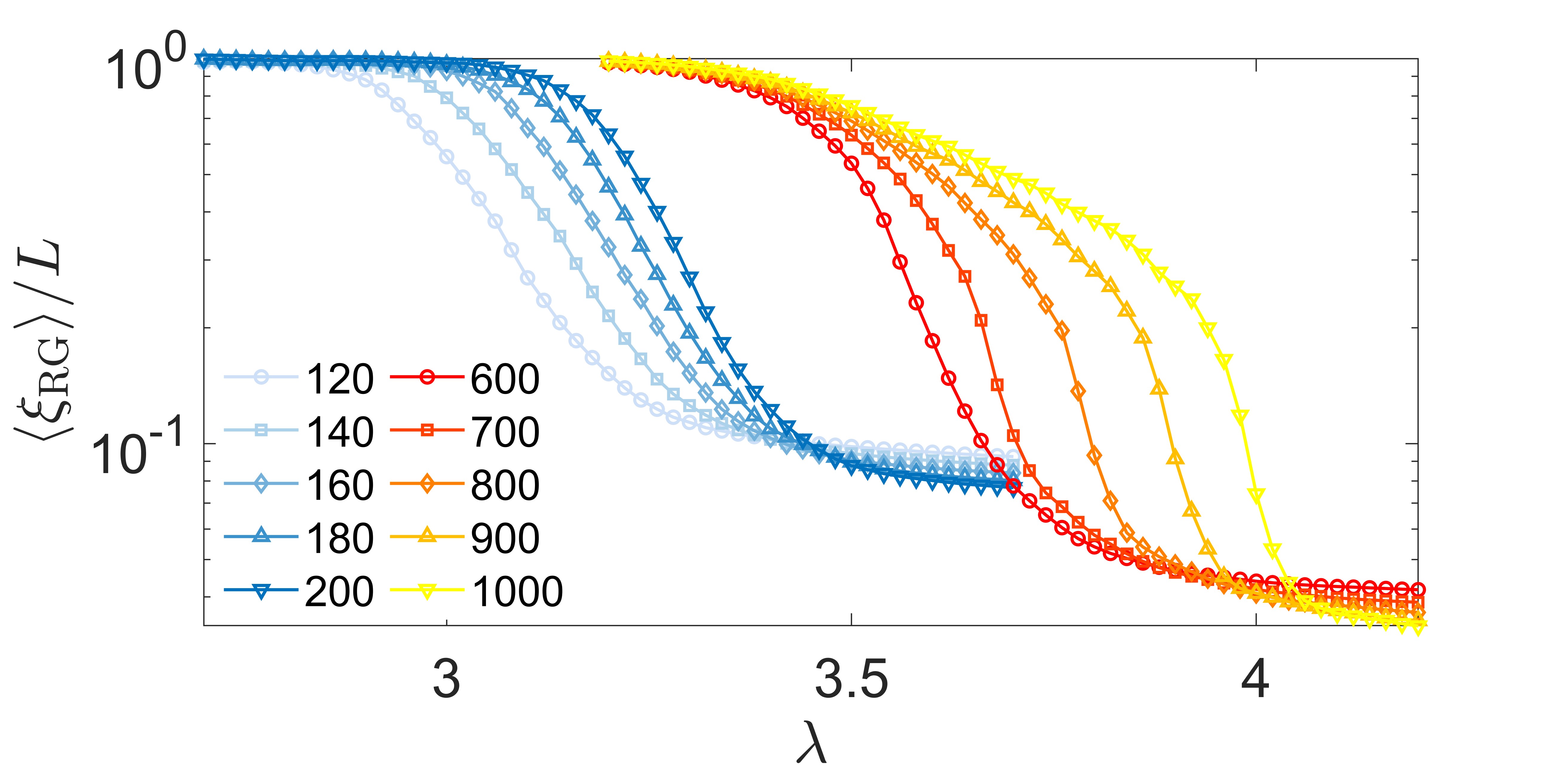}

\caption{\justifying {Critical properties of the MBL transitions for different system sizes. The parameters are $\alpha=1$ and $s=0.52$. The curves are averaged over $\sim10^5$ initial phases. Largest $L$ values indicate an upward drift in the apparent critical disorder.}}
\label{RG_avalanche}
\end{figure}

Although the size of the largest thermal clusters diverges with the system size, the whole system remains localized since $\xi_{\rm loc}/L\sim L^{d_{\rm MBL}-1}\to 0$ in the thermodynamical limit in the MBL phase (assuming the MBL phase to be stable). 
In analogous to the single-particle scenario, the physical picture of the SV MBL struture is that the individual isolated large clusters appear periodically along the chain, becoming larger and larger from the left side to the right side of the chain, whereas the regions in between these clusters are MBL-like. 
This behavior is physically quite different from the corresponding scenarios in random and QP MBL situations.

Note that the normalized EE should scale with the same exponent in the MBL phase, since its definition is directly related to the structure of the thermal clusters. However, the fluctuation in the scaling exponent calculated from the EE is larger, since the cut is not always located in the middle of the large clusters when averaging over different initial phases. This difference is not of any qualitative significance.

\subsection{Validity of RG and instability of MBL phase}

Given the consistency in the spatial structure in the single-particle localized eigentates as well as in the many-body resonant clusters, we first phenomenologically discuss what happens for parameter ranges away from the localized clusters, before exploring the critical properties of the MBL transition. 

First, for a given system size, the RG is only valid when $s$ is not too close to $1$ due to the initialization method, where it is assumed that the particles are localized around the charge center. 
This assumption is consistent with the asymptotic solution of the single-particle eigenstates, where the particles are localized in one large potential well. For a given system size, by increasing $s$, nearby potential wells come closer to each other and the particles can tunnel into nearby wells. 
In fact, we do see a failure of the RG procedure in Fig.~\ref{RG_fail}. 
Specifically, when $s=0.64$ (Fig.~\ref{RG_fail}(a)), the localization length at $L=600$ in the MBL phase fails to converge to the subvolume law. However, we recover the subvolume law using a larger system size with $L\geq 800$, although the fluctuation of the scaling exponent $d_{\rm MBL}$ is large compared with smaller $s$ as shown in Fig.~\ref{d_MBL_s}. When $s=0.7$ [Fig.~\ref{RG_fail}(b)], the RG completely fails at the system sizes we considered. 
This failure of the RG is also reflected by EE at $s=0.62$ [Fig.~\ref{RG_fail}(c)], but we believe the RG would become valid for $L>1000$, which is beyond our computational constraints.

Second, when $s$ decreases for a fixed system size, due to the increasing size of the local thermal clusters, the MBL phase could become unstable. 
In Fig.~\ref{RG_avalanche}, the curves of the localization length putatively intersect around one particular point, which is considered as the critical point $\lambda_c$, for small system sizes ($L=120$-$200$), 
which are, however, still much larger than the sizes accessible to the ED technique ($L<30$). 
However, the critical point becomes unstable and starts to drift to larger values of $\lambda$ for larger system sizes ($L=600$-$1000$). 
This indicates a critical length scale, above which the MBL phase may be unstable. We point out that this phenomenon is reminiscent of the avalanche instability~\cite{thiery2018many,morningstar2019renormalization,morningstar2022avalanches,sels2022bath,tu2023avalanche}, where the appearance of large thermal regions with the increasing system size could thermalize the whole system, although our RG here does not take into account the avalanche effect explicitly~\cite{dumitrescu2017scaling,zhang2018universal}. 
In fact, the slow upward drift of the critical disorder strength necessary to induce MBL at a fixed interaction strength with increasing system size also happens in the ED studies for random and QP models, 
and our RG analysis makes such a drift of the critical potential strength with system size in the SV model is consistent with these findings as well as with the avalanche studies.
The definitive analysis of the stability of the MBL in the SV model itself is beyond the scope of the current work.

\subsection{Critical properties and universalities}
\begin{figure}
\centering
\includegraphics[width=1\linewidth]{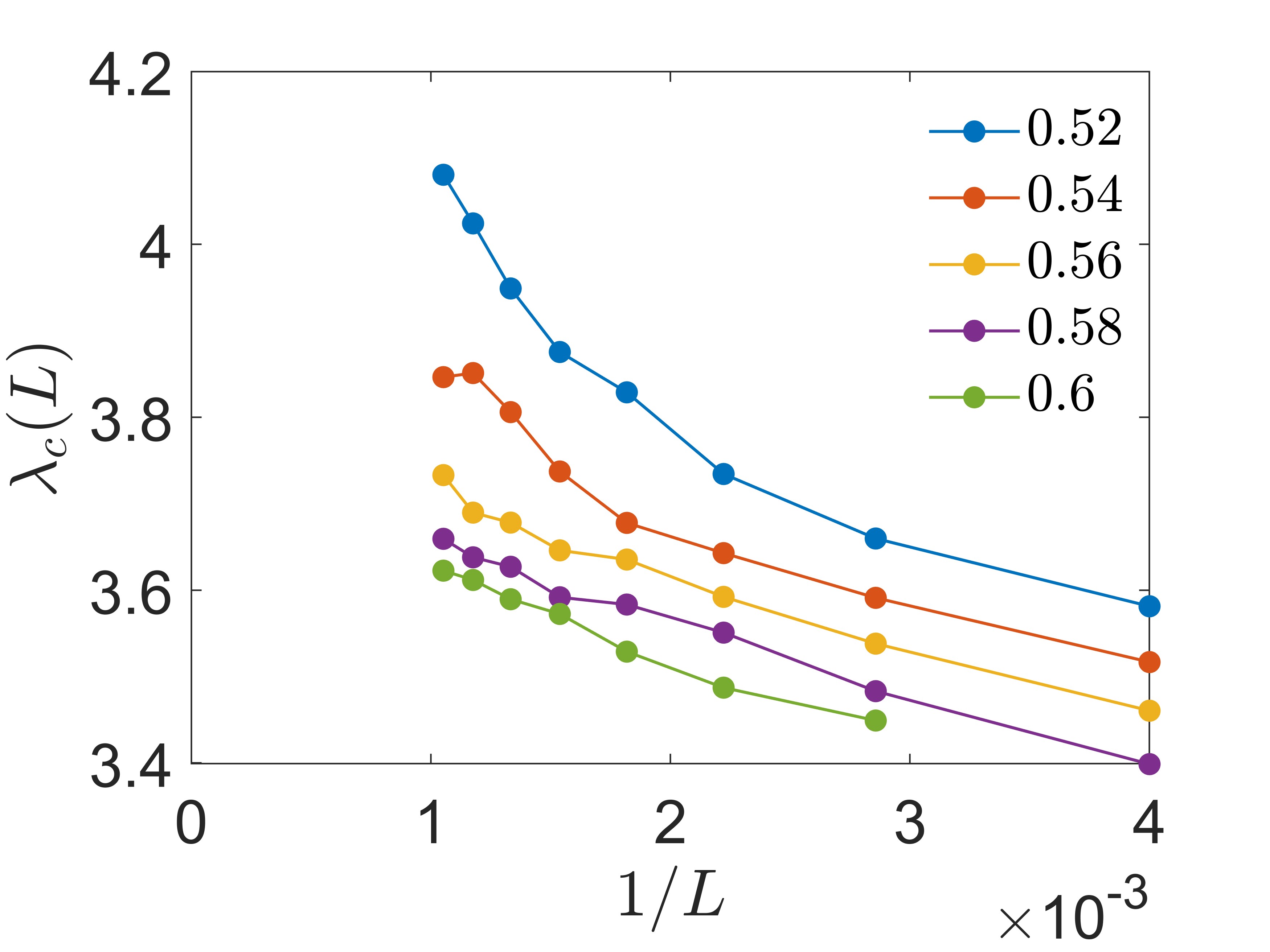}

\caption{\justifying { Drift of the critical disorder strength $\lambda_c$ as a function of system sizes $L$. Different colored lines show the systems with different slowly varying exponents. The $\lambda_c$ at $L$ is estimated as the crossing point between two scaling curves with $ L\pm50$ using the RG method. }}
\label{lmc_drift}
\end{figure}

\label{critical}

\begin{figure*}
\centering
\includegraphics[width=1\linewidth]{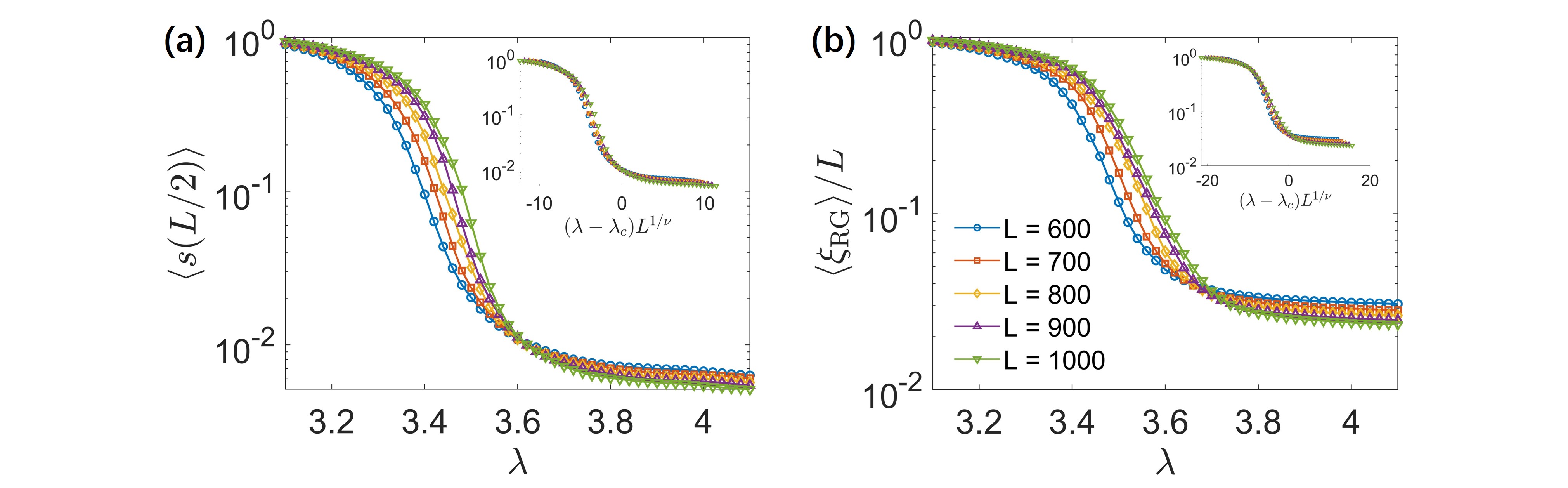}

\caption{\justifying {Slowly varying MBL transitions characterized by (a) the normalized entanglement entropy and (b) the localization length (defined as the length of the largest thermal clusters). For all figures, $\alpha=1,s=0.56$. The curves are averaged over $10^5\sim 10^6$ initial phases. Insets show the collapse of the data as a scaling function $f[(\lambda-\lambda_c)L^{1/\nu}]$.}}
\label{RG}
\end{figure*}

\begin{figure}
\centering
\includegraphics[width=1\linewidth]{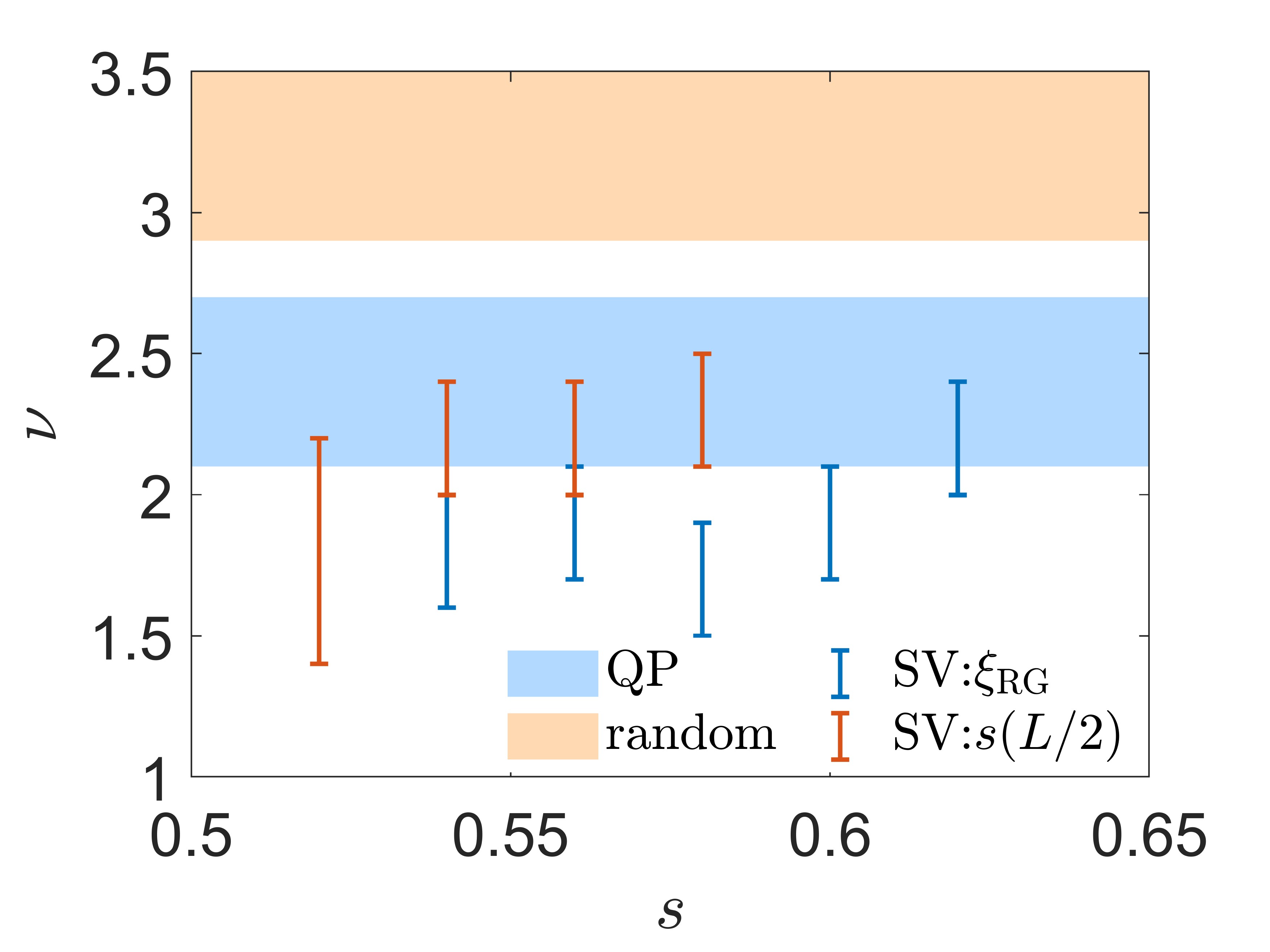}

\caption{\justifying {Calculated critical exponents for different slowly varying parameters $s$. The critical exponents are obtained using the scaling collapse methods for system sizes within $600$ and $1000$. Error bars with different colors denote the critical exponents obtained from different quantities. The colored areas denote the value of critical exponents for random ($\nu=3.5\pm0.3$)~\cite{dumitrescu2017scaling} and quasi periodic models ($\nu=2.4\pm0.3$)~\cite{zhang2018universal}. While the slowly varying (this work) and quasi periodic critical exponents mostly overlap within the error bars, the random model MBL exponent is higher beyond the error bars.}}
\label{nu_cri}
\end{figure}

Although the MBL phase could be unstable at a critical (and very large) length scale, below such a large length scale, however, the MBL transition also effectively flows to a fixed point and manifests a universal scaling behavior, analogous to probing a quantum phase transition at a finite temperature. To show this, we computed the drift of the critical disorder with increasing system sizes in Fig.~\ref{lmc_drift} with different $s$. Specifically, we estimate the $\lambda_c$ at $L$ as the crossing point between two curves as in Fig.~\ref{RG_avalanche} with $ L\pm50$. For $s \lesssim 0.54$, the drift gets faster for larger system sizes. When extrapolating $1/L$ to zero, the $\lambda_c$ likely diverges into infinity. This implies the MBL phase in the thermodynamic limit may not exist. In contrast, for $s \gtrsim 0.54$, the drift of $\lambda_c$ is similar to a linear function of $1/L$. If this holds to infinity, the MBL phase is stable. However, since we cannot rule out the possibility of faster drift for larger system sizes, the existence of the MBL phase in the thermodynamic limit remains an open question. Therefore, the criticality for the MBL transition is effective, which may or may not hold for the infinite system. 

To study the effective criticality of the SV MBL transition, we compute the normalized EE and localization length in a computationally feasible range of $s$ with system sizes up to $L=1000$. 
In Fig.~\ref{RG}, we see a clear MBL transition characterized by the normalized EE $s(L/2)$ and the localization length $\xi_{\rm RG}$ as a function of potential strength $\lambda$ for different system sizes $L$. 
Both quantities have a universal scaling form $f[(\lambda - \lambda_c)L^{1/\nu}]$ near criticality (inset of Fig.~\ref{RG}), where $\nu$ is the critical exponent. The critical quantities obtained from $s(L/2)$ and $\xi_{\rm loc}$ are $\lambda_c = 3.62\pm0.01,\nu = 2.2\pm0.2$ and $\lambda_c = 3.68\pm0.02,\nu = 1.9\pm0.2$. The critical exponents agree with each other within error bars. 

To confirm that critical exponent ($\nu\approx 2$) indeed characterizes the universality class for the MBL transition, we calculate the critical exponent using different SV exponent $s$ values as shown in Fig.~\ref{nu_cri}. 
We find that within the parameter regime, the critical exponent remains approximately a constant. 
This gives confidence that the SV MBL transition indeed belongs to a universality class with a unique critical exponent which is independent of the exponent controlling the SV on-site potential inducing the MBL in the interacting system. 
We compare the critical exponent of the SV model with other models.
Remarkably, the critical exponent of the SV model differs from the random model where $\nu \gtrsim 3$~\cite{dumitrescu2017scaling,vosk2015theory,potter2015universal} and is quite close to the QP model $\nu = 2.4 \pm 0.3$~\cite{zhang2018universal}, but is still $10\%$ below the QP exponent.
This implies that the SV MBL and the QP MBL transition may belong to the same universality class unless the small difference ($\sim 10\%$) between the two exponents is construed to be significant.
More work is necessary to decisively settle whether SV and QP models belong to the same or different universality classes because of the rather small difference in their exponents, but it is reasonably certain that the random MBL universality, with an exponent which is almost $40\%$ larger, is different. 
Although it is not clear whether all deterministic models, such as QP and SV model, with the MBL transition belong to the same universality class, current results implies in the existence of at least two different universality classes of the MBL transition, which are governed by the infinite randomness and the non-random fixed points, respectively~\cite{khemani2017two}.

\section{Conclusion}
\label{section_Con}

We have studied the MBL transition of an interacting model with an SV deterministic potential using both ED and RG. For small system sizes, such an MBL transition is qualitatively similar to that of the other models, with a transition of the EE and the level statistics indicating MBL. Also, a logarithmic growth of EE from weakly entangled states is found in the MBL phase. In addition, we confirm the MBL transition using the real-space RG method in the asymptotic regime. Different from conventional MBL models, a subvolume scaling of EE is observed in the MBL phase, as a consequence of the emergence of arbitrarily large clusters which scale with the system size. In the regime where the MBL transition is stable, the EE and the localization length have a universal scaling form with a critical exponent $\nu\approx 2$ which is independent of the quantitative details of the SV potential. Since this scaling exponent is close to that of the QP model, we believe that SV and QP MBL transitions belong to the same universality class, which differs from the MBL in the random disorder case. We cannot, however, completely rule out the possibility that SV and QP models also have distinct universality classes since their critical MBL exponents differ by 10\%, but more numerical work on larger systems would be necessary to settle this issue.

We point out that quantum criticality in MBL is not established -- our work only suggests an effective criticality in RG system sizes of $1000$, which is two orders of magnitude larger size than what was originally used in asserting that MBL exists as a dynamical phase transition~\cite{pal2010manybody}. 
The fact that the critical disorder slowly increases with increasing system size clearly indicates that no existing theory, including ours, can assert the existence of MBL since the asymptotic thermodynamic limit has not been reached in spite of 15 years of concerted research efforts.  Either MBL does not exist at any finite disorder (i.e., no criticality) in the thermodynamic limit, or it exists for very large disorder in very large systems which numerics cannot reach. This conclusion is consistent with various avalanche analyses of MBL in random~\cite{morningstar2019renormalization,morningstar2022avalanches} and QP\cite{tu2023avalanche} models.  We note that Imbrie's mathematical proof for MBL \cite{imbrie2016many} involves subtle assumptions at the starting point, which may or may not be satisfied by the physical systems.  
More importantly, Imbrie's proof is explicitly constructed for random systems and does not apply to QP or SV deterministic models. So, Imbrie's proof is irrelevant for our MBL studies in SV potential. Therefore, the question of whether MBL exists or not in any system in the thermodynamic limit remains open although MBL certainly exists as an effective phase transition up to very large (perhaps even experimentally relevant) system sizes. The fact that MBL seems more robust in QP models~\cite{tu2023avalanche} indicates that we cannot appeal to Imbrie's proof as the justification for the existence of MBL in the thermodynamic limit. In this context, it is useful to also mention that the effective finite-size MBL persists to accessibly large system sizes even in models with SPMEs with an effective critical disorder strength not that different from that in models without SPMEs\cite{Li2015manybody,li2016quantum} (see our Appendix \ref{apdxB}). Our current work should therefore be considered as studying the effective quantum criticality in MBL, just as every MBL study in the literature has been done over the last two decades. The question of MBL in the thermodynamic limit remains open and elusive.

Several interesting aspects of the model should be investigated in the future. One question is the stability of the SV MBL. In other MBL models, it has been recently discovered that the MBL phase may be unstable due to the avalanche instability~\cite{thiery2018many,morningstar2022avalanches,sels2022bath}, where large thermalizing regions could thermalize the whole system up to very large disorder strength. Since such a mechanism is not considered explicitly in our RG analysis, it is not clear whether an actual avalanche transition point \cite{morningstar2019renormalization,tu2023avalanche} exists in the SV MBL. 
We do, however, see suggestive hints of MBL instability in our RG work for our largest system sizes, and speculate that the SV model also suffers from MBL instabilities demonstrated in the random and QP models.  
It is therefore possible to study the avalanche instability of MBL in a smaller system size, which may be captured by the experiments explicitly.
This physics requires more work in the future.

We mention in this context that the single-particle SV model has a mobility edge~\cite{sarma1988mobility,sarma1990localization}, but we have used potential parameters in the SV potential for all our work here ensuring that the non-interacting system is always in the completely localized phase, thus avoiding all questions about the possible existence of any intermediate non-ergodic extended many-body phase much discussed in the literature for various duality-breaking QP models~\cite{Li2015manybody,modak2015manybody,li2016quantum,kohlert2019observation,deng2017many,huang2023incommensurate,tu2023localization,tu2024interacting}. Another possible future work could focus on the role of the SPME in MBL in the interacting SV potential model. We notice that Ref.~\cite{nag2017manybody} discusses the many-body mobility edge in the interacting SV model with ED, but the further investigation on the non-ergodic extended regime is not mentioned.

\section*{Acknowledgements}
The authors thank D.\ Vu and D.\ M.\ Long for useful discussions.
This work is supported by the Laboratory for Physical Sciences. The authors acknowledge the University of Maryland supercomputing resources (\href{https://hpcc.umd.edu}{https://hpcc.umd.edu}) made available for conducting the research reported in this paper.

\appendix

\section{Solutions for single-particle eigenstates and the mobility edges} \label{apdxA}
Following the analysis in Ref.~\cite{sarma1990localization}, we review the analytical solution for the single-particle eigenstates of the SV model. Consider a single-particle eigenstate $|\psi\rangle = \sum_j \psi_j |j\rangle$, where $\psi_j$ is the amplitude on each lattice site and satisfies
\begin{equation}
    \psi_{j+1} + \psi_{j-1} + V_j \psi_{j} = E \psi_{j}
\end{equation}
where $V_j = \lambda \cos(\pi \alpha j^s + \phi)$ and $E$ is the eigenvalue. Let $j = N+l$ and $N\gg l$ is a constant, we have 
\begin{equation}
\begin{aligned}
    V_{N+l} \approx & \lambda \cos(\pi \alpha s N^{s-1}l + \pi \alpha s N^{s} + \phi) \\ \equiv & \lambda \cos(\pi \alpha s N^{s-1}l + \phi_N)
\end{aligned}
\end{equation}
Define $x \equiv N^{s-1}l$, which can be considered as a continuous variable, and the equation becomes
\begin{equation}
    \psi(x+N^{s-1}) + \psi(x-N^{s-1}) = [E - \cos(\pi \alpha s x + \phi_N)] \psi(x)
\end{equation}
Now we can apply the WKB approximation by assuming the solutions to be $\psi(x) = e^{-f_0(x)N^{1-s}}$. Expand $f_0(x\pm N^{s-1}) = f_0(x) \pm N^{s-1} f_0'(x)$, we have
\begin{equation}
    f_0'(x) = \cosh^{-1}\left[\frac{E - \lambda\cos(\pi \alpha s x + \phi_N)}{2}\right]
\end{equation}
Therefore, 
\begin{equation}
    f_0(x) = \int_0^x \cosh^{-1}\left[\frac{E - \lambda\cos(\pi \alpha s x' + \phi_N)}{2}\right] dx'
\end{equation}
Let $\theta = \pi \alpha s x' + \phi_N$, consider the average localization length within one potential period
\begin{equation}
    {\xi}^{-1} = \frac{1}{2\pi} \int_0^{2\pi} {\rm Re} \cosh^{-1} \left[\frac{E - \lambda\cos\theta}{2}\right] d\theta
\end{equation}
If the eigenstate is localized, $\bar{\xi}^{-1} > 0$, which requires
\begin{equation}
    \sup |E - \lambda\cos\theta|>2
\end{equation}
Such a condition cannot be satisfied if $\lambda<2$ and $-2+\lambda<E<2-\lambda$, where the states are fully extended. Thus, the mobility edge exists at $E_c = \pm(2-\lambda)$.

We mention that the non-zero average localization length does not imply the eigenstate is localized at all sites. No matter what the potential strength $\lambda$ is, we always find a region where ${\rm Re}\cosh^{-1}[E - \lambda\cos\theta] = 0$. For example, in the middle of the spectrum $E=0$, in the region $\cos^{-1}(2/\lambda)<\theta<\pi - \cos^{-1}(2/\lambda)$, eigenstates are extended. This characterizes the size of the locally extended regions $\sim N^{1-s}/(s\alpha)$ within each potential period, as mentioned in the main text.

For all the MBL work presented in the main text of this paper (except for those in Appendix \ref{apdxB}), we use a potential strength $\lambda>2$, thus implying that all the single-particle states are localized with no mobility edge physics affecting the single-particle spectrum.

\section{Additional results on the interacting SV model}
\label{apdxB}
In this section, we study other aspects of the interacting SV model using ED. In particular, we compute the interacting phase diagram and compare it to that of the AA model. We also compute the energy-resolved phase diagram to explore the many-body mobility edge. Finally, we discuss the relation between the EE and the on-site potential in different parameter regimes to explore any possible non-ergodic extended behavior of the model arising from the existence of the SPME in the single-particle spectrum.

\subsection{Interacting phase diagram}
To identify the interacting phase diagram of the SV model, we evaluate two quantities: the mean gap ratio and the many-body inverse participation ratio (MIPR). The mean gap ratio follows the definition in the main text. The MIPR 
\begin{equation}
    \mathcal{I} = \frac{1}{1-f}\left[\frac{1}{fL}\sum_{j=1}^L \langle n_j\rangle^2 - f\right]
\end{equation}
is used to characterize the localization of the many-body eigenstates, where $\langle n_j\rangle$ is the average particle number over the many-body eigenstates $|\psi\rangle$ on site $j$. 
We choose the filling fraction $f=1/6$ as in the main text. The MIPR describes the spatial distribution of the eigenstates. Namely, the eigenstates are fully localized if $\mathcal{I}\rightarrow 1$, and fully extended if $\mathcal{I}\rightarrow 0$. The MIPR averaged over all many-body eigenstates $\langle \mathcal{I}\rangle$ is used to characterize the phase diagram.

As a comparison, we also compute these quantities of the interacting AA model, whose non-interacting Hamiltonian does not have any SPME, and for $\lambda>2$ its eigenstates are all localized. (By contrast, of course, the SV model does have the SPME.) In our context, the AA model corresponds to setting $s=1$ and $\alpha$ to be irrational in the SV model. In particular, we choose $\alpha = 1+\sqrt{5}$ throughout. 

\begin{figure*}
\centering
\includegraphics[width=.8\linewidth]{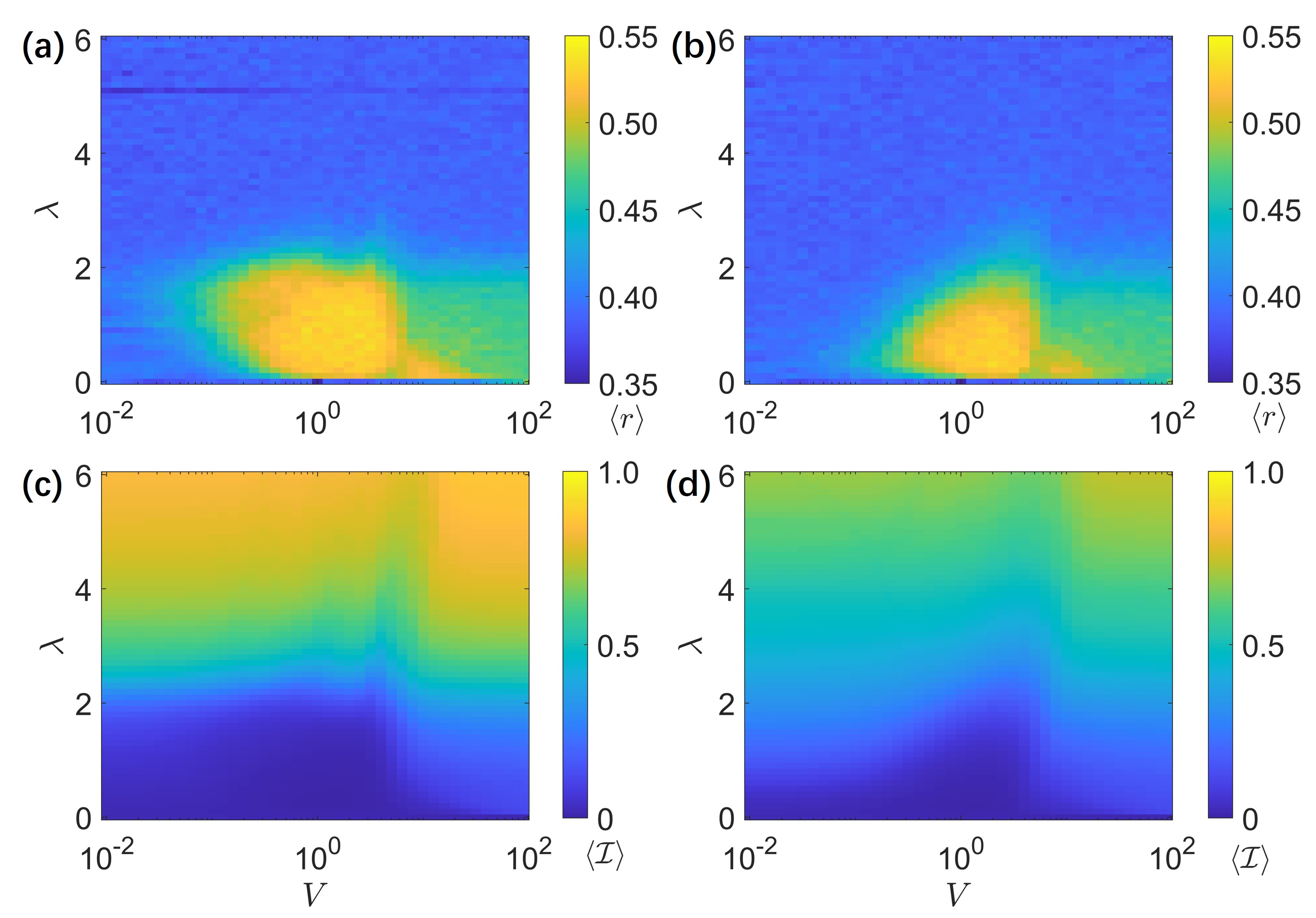}

\caption{\justifying {The interacting ED phase diagram: [(a) and (c)] the Aubry-Andr\'e model with $\alpha=1+\sqrt{5}$ and $s=1$ and [(b) and (d)] the SV model with $\alpha=1$ and $s=0.7$ characterized by the mean gap ratio $\langle r \rangle$ and the averaged many-body inverse participation ratio $\langle \mathcal{I}\rangle $. For all figures, we set $L=24$ with filling fraction $f=1/6$. }}
\label{IPD}
\end{figure*}

In Fig.~\ref{IPD}, we calculated the phase diagrams of the AA model and the SV model using the mean gap ratio and the MIPR. Similar to other models \cite{huang2023incommensurate}, the phase diagram can be divided into three regimes according to the different behaviors of the model with different interactions -- the weakly interacting regime ($V<V_1$), the intermediate regime ($V_1<V<V_2$) and the strongly interacting regime ($V>V_2$), where $V_1\sim 0.1$ and $V_2\sim10$. 
In the weakly interacting regime, the many-body eigenstates are close to the product of single-particle eigenstates if the interaction is considered as a first-order perturbation. 
This is supported by the numerically calculated MIPR in the weakly interacting regime. For the AA model, a delocalized-localized transition appears near $\lambda \sim 2$. While for the SV model, there is no such clear transition due to the presence of the SPME. The weak interaction fails to thermalize the system as shown in Figs.~\ref{IPD}(a) and \ref{IPD}(c). 

In the intermediate regime, both models have a thermal to non-thermal phase transition in Figs.~\ref{IPD}(a) and \ref{IPD}(b). Notice that such transitions appear along with the localization-delocalization transition in Figs.~\ref{IPD}(c) and \ref{IPD}(d). 
Both transitions are signatures of MBL, as we pointed out in the main text. Another interesting observation is the difference in the critical potential strengths $\lambda_c$ distinguishing the MBL-to-ergodic transitions. In the AA model, $\lambda_c$ is almost independent of the interaction strength, while $\lambda_c$ increases with interaction strength in the SV model. This is possibly due to the presence of the SPME in the SV model, where the thermalization of localized states requires larger interactions. 
We mention that, as a matter of principle, finite-size effects also play a crucial role in determining the critical disorder strength for the MBL transition, but this dependence is very slow and difficult to discern in ED calculations due to the exponential growth of the Hilbert space with the increasing system size.

In the strongly interacting regime, since the on-site potential can be considered as a perturbation, the two models become similar. We notice that the mean gap ratio for $\lambda<2$ is neither Poissonian nor Wigner-Dyson type. This phenomenon can be understood as follows. In the strongly interacting limit, the whole Hilbert space is fragmented into dynamically disconnected subspaces according to the massive energy difference of different particle configurations \cite{vu2022fermionic,huang2023incommensurate}. In particular, for a two-particle case, the subspaces are divided by whether two particles are adjacent to each other or not. The mean gap ratio between different subspaces is Possonian, and within each subspace, it is ergodic. Therefore, the mean gap ratio is still Wigner-Dyson type. The mean gap ratio is thus the summation of these two types. Because of our choice of the filling fraction at a particular system size, the summation of the gap ratio is between $0.38$ and $0.53$. 

We note that in the localized phase, the eigenstates of the SV model are less localized than those of the AA model according to the MIPR, which is possibly due to the presence of the local thermal regions as discussed in the main text.

\subsection{Energy-resolved phase diagram}
\begin{figure*}
\centering
\includegraphics[width=.8\linewidth]{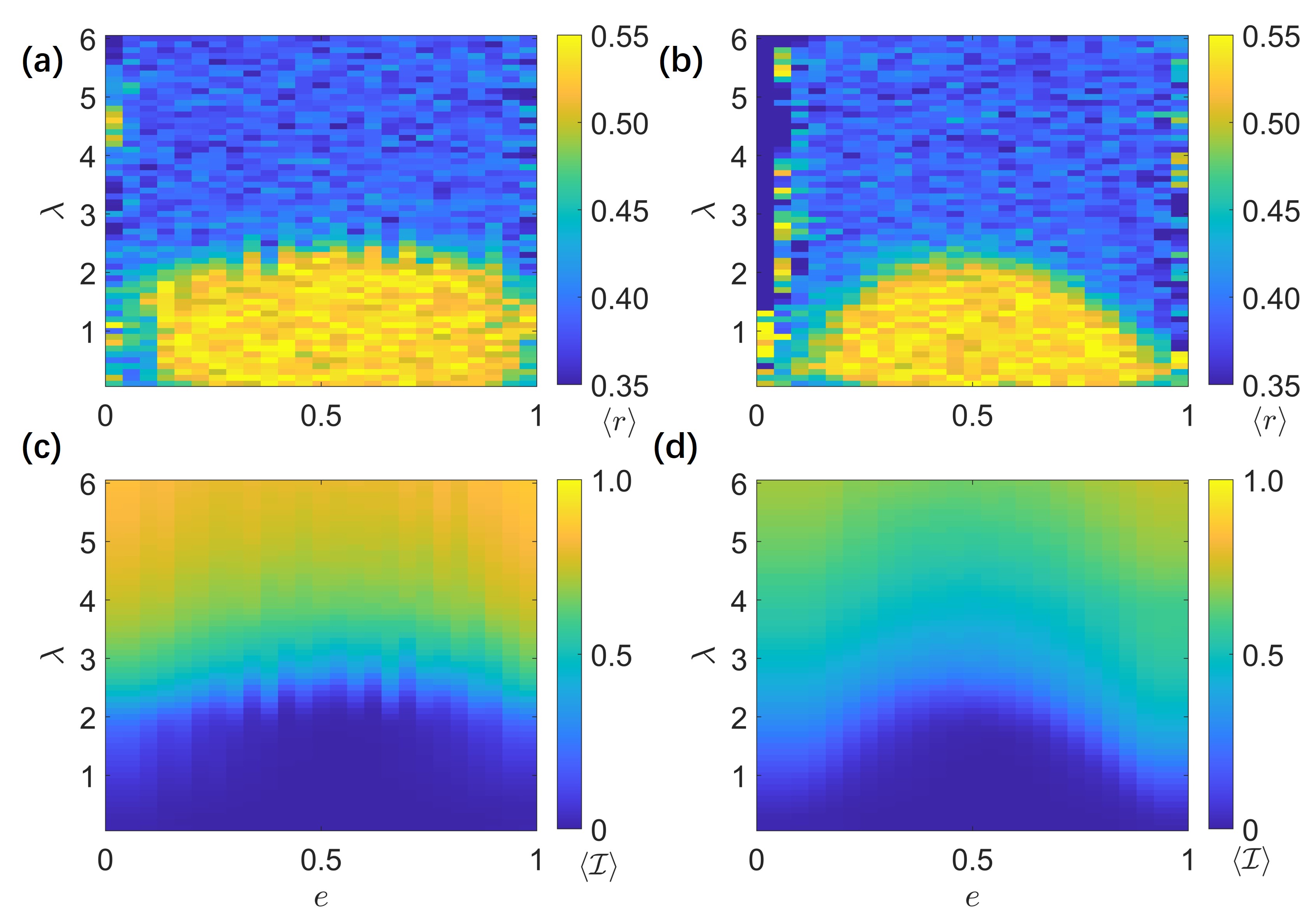}

\caption{\justifying {The energy-resolved interacting ED phase diagram: [(a) and (c)] the Aubry-Andr\'e model with $\alpha=1+\sqrt{5}$ and $s=1$ and [(b) and (d)] the SV model with $\alpha=1$ and $s=0.7$ characterized by the mean gap ratio $\langle r \rangle$ and the averaged many-body inverse participation ratio $\langle \mathcal{I}\rangle $. For all figures, we set $L=30$ and $f=1/6$. We average over 200 eigenstates in the middle of each energy interval. }}
\label{EPD}
\end{figure*}

To explore the possibility of many-body mobility edges (MBME) \cite{Li2015manybody} arising from the SPME, we also show the energy-resolved phase diagrams at a fixed interaction strength $V=1$ using the same quantities in Fig.~\ref{EPD}, where $e=(E-E_{\rm min})/(E_{\rm max}-E_{\rm min})$ is the normalized energy. For both models, we again observe an MBL-to-ergodic transition from small to large potential strength characterized by the mean gap ratio and the MIPR. We notice that the energy dependence of the critical potential strength $\lambda_c(E)$ (MBME) is slightly different for the two models. 
However, because of the finite-size limitations of the ED, this observation can hardly be interpreted as a qualitative difference of the MBME as a consequence of the SPME.
We are not making any rigorous arguments; instead, just providing some insights for the possible MBME in the model using numerical ED results and heuristic perturbative arguments.

\subsection{Relation between entanglement entropy and local potentials}
\label{EELP}
To provide some intuition on the difference in the MBL phases between the deterministic and random models, we study the differential EE
\begin{equation}
    \Delta S(l) = S(l+1) - S(l),
\end{equation}
where $S(l)$ is the EE of the subsystems $l$, as a reflection of LIOM structure in the MBL phase. Specifically, if one of the (or both) cuts is located within a local thermal bubble, then we have $\Delta S(l) = \pm s_{\rm th} = \pm\ln 2$ (the $\pm$ comes from the position of the cut in the thermal bubble). 
On the other hand, if both cuts are not in the bubble, then $\Delta S(l) = 0$. In addition, this quantity is also useful for detecting the MBL-thermal transition by averaging over disorders or initial phases. 
In the thermal phase, the probability of cutting into a thermal region approaches one. Therefore, $\langle\Delta S(l)\rangle=s_{\rm th}$. In the MBL phase, although the probability of cutting into a thermal region is non-zero, we have equal probability of cutting into the left and right sides of the cluster, therefore $\langle\Delta S(l)\rangle=0$.

\begin{figure*}
\centering
\includegraphics[width=1\linewidth]{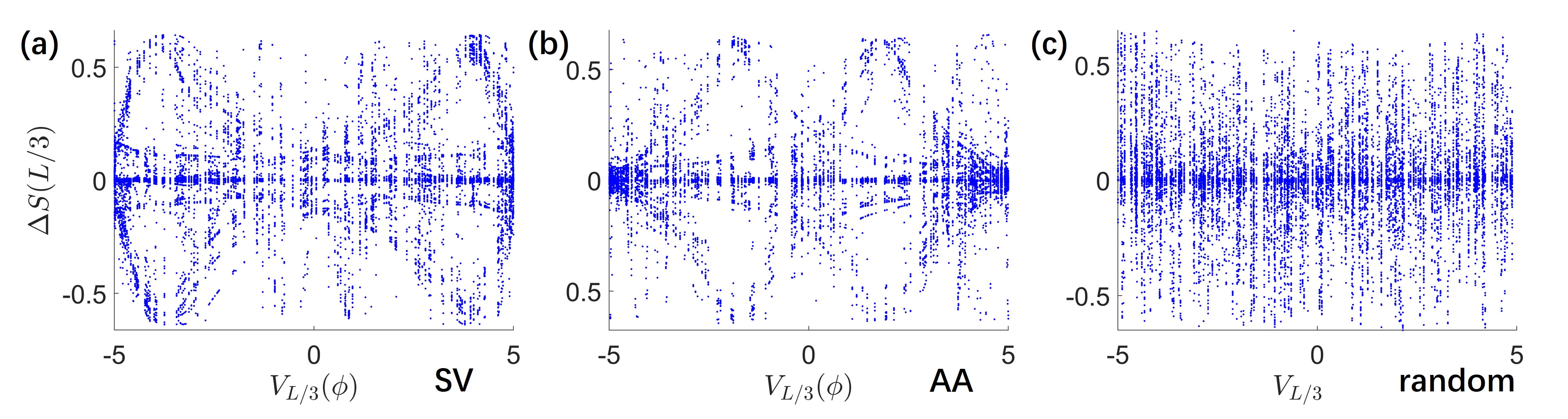}

\caption{\justifying {The relation between differential entanglement entropy $\Delta S(l) = S(l+1)-S(l)$ (we choose $l=L/2$) and the on-site potentials of (a) slowly varying model, (b) Aubry-Andr\'e model, (c) random model. For all figures, $\lambda=1.5,L=24,\alpha=1.1,s=0.71$. The local potentials are tuned by initial phases $\phi$ and the disorders in the deterministic and random models, respectively. We compute the $50$ eigenstates in the middle of the spectrum for a particular initial phase or a disorder realization. }}
\label{EE_phi_models}
\end{figure*}

In particular, we study the relation between the differential EE $\Delta S(l)$ of the multiple eigenstates within a specific energy interval and the potential at the same site $l$. We choose $\lambda=5$ for different models so that the system is in MBL phase, with $\langle\Delta S(l)\rangle\approx0$ in Fig.~\ref{EE_phi_models}.
Moreover, the particular shape of differential EE indicates that the local thermal bubbles in the SV and AA models are not randomly distributed along the chain but have a particular relation with the on-site potentials. Such a property is not seen in the random model, where the on-site potentials are randomly distributed within $[-\lambda,\lambda]$. 
In the main text, we have also made a similar argument using RG for the SV model, the thermal regions in the MBL phases are distributed around the local minimum and maximum of the potentials, which is also reflected by the results in Fig.~\ref{EE_phi_models}(a). 

\begin{figure*}
\centering
\includegraphics[width=1\linewidth]{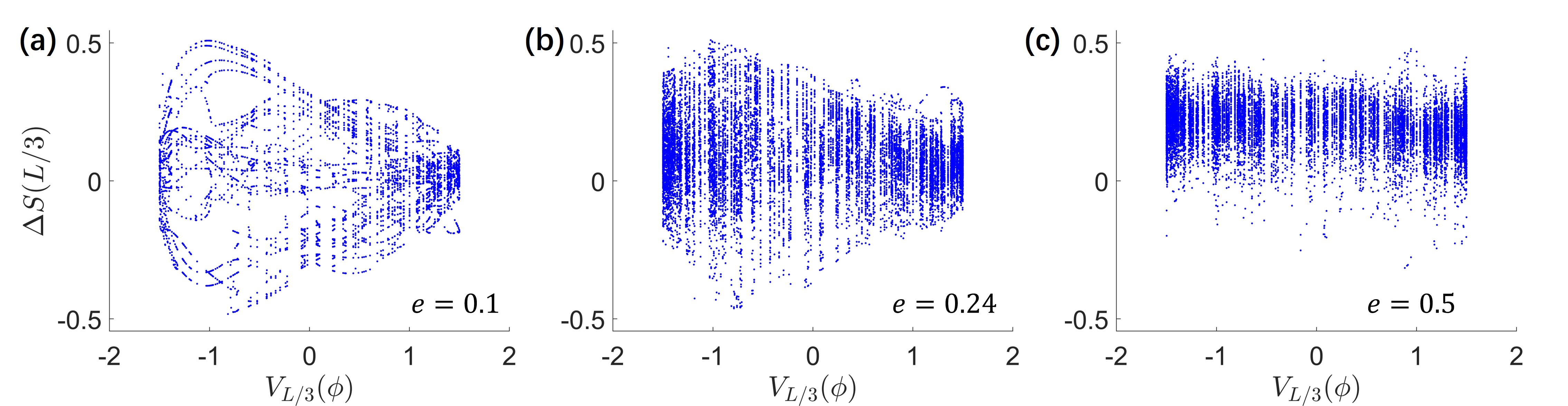}

\caption{\justifying {The relation between differential entanglement entropy $\Delta S(l) = S(l+1)-S(l)$ (we choose $l=L/2$) and the on-site slowly varying potentials at different energy. For all figures, $\lambda=1.5,L=24,\alpha=1.1,s=0.71$. For the systems with different initial phase $\phi$, we compute the $50$ eigenstates around the relative energy point $e$. }}
\label{EE_phi_SV}
\end{figure*}

We mention that the relation between differential EE and the local potential in deterministic systems whose ensembles are parameterized by a finite number of parameters (e.g.\ the initial phase $\phi$ in the SV and AA models), may reflect deeper properties of the MBL transitions.
In our case, the local potential can be directly associated with the initial phase of the potential function, and a small change in it corresponds to a small change in the system.
In Fig.~\ref{EE_phi_SV}, we show the differential EE of the SV model with $\lambda=1.5$ at different positions on the spectrum.
In the localized regime (Fig.~\ref{EE_phi_SV}(a)), one can observe that the dots form multiple orbit-like structures.
This can be understood as due to the small sizes of the LIOMs, the value of $\Delta S$ only depends on a few of them near the cut. 
A small change of the parameter then corresponds to a perturbation to the local system, which modifies the shape of the LIOMs only a little bit, thus forming continuous curves.
Different curves correspond to different eigenvalues of the LIOMs, and theoretically there can be infinitely many curves very close to each other in a given energy window, as re-configuring the eigenvalues of asymptotically faraway LIOMs can lead to similar energy but with asymptotically small change in $\Delta S$.
As the system delocalizes, the infinite set of curves smears out more and more [(]Fig.~\ref{EE_phi_SV}(b)], possibly due to the lengthening of the LIOMs and the instability of the LIOM shape against the perturbation of the parameters. Finally, in the thermal phase [(]Fig.~\ref{EE_phi_SV}(c)], there is no such orbital structure at all.

Probing the thermalization or delocalization transition based on this observation, however, is not easy.
As in other ED methods, it suffers from finite-size effects. In addition, to observe the orbit structure near the transition point, one needs a huge number of sampling points.
Theoretically, it is also not clear when and how the orbit-like structure breaks down, as the notion of adiabaticity (which would give a clear definition of ``orbits'' even if nearby orbits already smear out) is unclear in such systems.
Therefore, we do not go further into this direction, and leave this observation for future studies.
We mention that in the asymptotic regime of the SV system, tuning the initial phase mimics the Thouless pump in some sense~\cite{PhysRevB.27.6083,Citro2023}.

\bibliographystyle{apsrev4-1}
\bibliography{ref}

\end{document}